\documentclass[twocolumn,prl,floatfix,citeautoscript,nofootinbib,superscriptaddress]{revtex4}
\usepackage{amsbsy}
\usepackage{latexsym,epsfig,graphicx}
\usepackage{dcolumn}
\usepackage{graphicx}
\usepackage{subfigure}
\usepackage{comment}
\usepackage{color}
\usepackage{bm}
\usepackage{mathrsfs}
\usepackage{amsfonts}
\usepackage{amsmath}
\usepackage{color}
\usepackage{amssymb}
\usepackage{xspace}
\usepackage{epstopdf}
\usepackage{tabularx}
\usepackage{longtable}
\usepackage[colorlinks=true, letterpaper=true, pdfstartview=FitV, linkcolor=red, citecolor=blue, urlcolor=blue]{hyperref}
\usepackage[normalem]{ulem}

\pdfoutput=1

\begin{document}

\title{Pseudo-Goldstone Excitations in a Striped Bose-Einstein Condensate}
\author{Guan-Qiang Li}
\affiliation{Department of Physics, The University of Texas at Dallas, Richardson, Texas
75080-3021, USA}
\affiliation{School of Arts and Sciences, Shaanxi University of Science and Technology,
710021 Xi'an, China}
\author{Xi-Wang Luo}
\thanks{Corresponding author email: xiwang.luo@utdallas.edu}
\affiliation{Department of Physics, The University of Texas at Dallas, Richardson, Texas
75080-3021, USA}
\author{Junpeng Hou}
\affiliation{Department of Physics, The University of Texas at Dallas, Richardson, Texas
75080-3021, USA}
\author{Chuanwei Zhang}
\thanks{Corresponding author email: chuanwei.zhang@utdallas.edu}
\affiliation{Department of Physics, The University of Texas at Dallas, Richardson, Texas
75080-3021, USA}

\begin{abstract}
Significant experimental progress has been made recently for observing
long-sought supersolid-like states in Bose-Einstein condensates, where
spatial translational symmetry is spontaneously broken by anisotropic
interactions to form a stripe order. Meanwhile, the superfluid stripe ground
state was also observed by applying a weak optical lattice that forces the
symmetry breaking. Despite of the similarity of the ground states, here we
show that these two symmetry breaking mechanisms can be distinguished by
their collective excitation spectra. In contrast to gapless Goldstone modes
of the \textit{spontaneous} stripe state, we propose that the excitation
spectra of the \textit{forced} stripe phase can provide direct experimental
evidence for the long-sought gapped pseudo-Goldstone modes. We characterize
the pseudo-Goldstone mode of such lattice-induced stripe phase through its
excitation spectrum and static structure factor. Our work may pave the way
for exploring spontaneous and forced/approximate symmetry breaking
mechanisms in different physical systems.
\end{abstract}

\maketitle

{\color{blue}\emph{Introduction.---}}Spontaneous symmetry breaking plays a
crucial role for the understanding of many important phenomena in different
fields ranging from elementary particles to condensed states of matter. For
instance, crystalline and superfluidity orders are formed in the long-sought
supersolids through spontaneously breaking spatial translational and U(1)
gauge symmetries~\cite{BoninsegniSupersolids2012}. While the early study of
supersolidity focused on solid $^{4}$He~\cite{thouless1969the,
andreev1971quantum} without conclusive experimental evidence \cite{He1,He2},
ultracold atoms have emerged as a powerful platform in recent years for
observing supersolid-like quantum phases~\cite{cinti2014defect,
baumann2010dicke, PhysRevLett.113.070404,
leonard2017supersolid,PhysRevLett.122.130405,
PhysRevX.9.011051,PhysRevX.9.021012,NataleExcitation2019,LiA2017}.
Significant experimental progress has been made on the generation and
measurement of supersolid-like superfluid stripe states in both dipolar~\cite%
{PhysRevLett.122.130405,PhysRevX.9.011051,PhysRevX.9.021012,NataleExcitation2019}
and spin-orbit-coupled Bose-Einstein condensates (BECs)~\cite{LiA2017},
where spontaneous translational symmetry breaking is driven by dipolar or
anisotropic spin interactions. In the latter case, the anisotropic spin
interactions favor the occupation of both band minima of the spin-orbit
coupling induced double-well dispersion, yielding superfluid stripe phase
with periodic density modulations~\cite%
{ChunjiWang2010,TLHo2011,Lin2011,YunLi2012,J.Sanchez-Baena2020,XWL2017}.

In the region where ground state symmetry cannot be spontaneously broken by
interactions, the symmetry breaking ground state may be achieved by applying
a weak symmetry breaking potential. Such forced symmetry breaking mechanism
has been demonstrated recently in a spin-orbit-coupled BEC, where the
superfluid stripe ground state is realized by applying a weak optical
lattice~\cite{BersanoExperimental2019} that breaks translational symmetry
explicitly. Interestingly, the forced stripe ground state shows similar
(spin-)density patterns as the spontaneous one induced solely by the
anisotropic atomic interactions. Therefore two questions naturally arise:
Can we distinguish the stripe ground states resulted from spontaneous and
forced symmetry breaking mechanisms? If so, are there interesting
experimental observables? These questions should also apply to general
spontaneous and forced symmetry breaking ground states in other physical
systems.

In this Letter, we address these two important questions by investigating
the collective excitations of the forced superfluid stripe ground state and
showing that the emerging pseudo-Goldstone spectrum lies at the heart of
understanding its forced symmetry breaking. The pseudo-Goldstone mode is an
important concept in fields ranging from standard model to solid-state
materials~\cite%
{GoldstoneBroken1962,BurgessGoldstone2000,WeinbergApproximate1972}, with
prominent examples including the pion (the lightest hardon) \cite%
{NambuAxial1960,NambuDynamical1961} and longitudinal polarization components
of $W$ and $Z$ bosons in high-energy physics~\cite{TanabashiReview2018},
phonon modes in superconductors and/or superfluids~\cite%
{NambuQuasi1960,HartmannDirect1996,DemlerSO2004,FernandesLow2017} and
magnons in magnets~\cite{ChumakMagnon2015,HenleyOrdering1989,RauPseudo2018}.
However, direct experimental observation of pseudo-Goldstone spectrum
remains challenging. The capability of directly measuring excitation
spectrum using Bragg spectroscopy~\cite%
{J.Stenger1999,J.Steinhauer2002,PhysRevA.90.063624,PhysRevLett.114.105301,R.Mottl2012,PhysRevLett.114.055301}
in ultracold atomic gases thus provides a powerful tool for probing
pseudo-Goldstone spectrum. Our main results are:

\textit{i)} In the strong anisotropic spin interaction region, the
spontaneous superfluid stripe ground state hosts two gapless Goldstone
modes. A weak lattice breaks the translational symmetry (i.e., the symmetry
is approximate) and turns one gapless mode into a pseudo-Goldstone mode,
which is characterized by the gap of the excitation spectrum at the long
wavelength limit (i.e., zero-momentum gap). The hybridization of the gapped
pseudo-Goldstone and the remaining gapless modes yields an avoided-crossing
gap at a finite momentum.

\textit{ii}) In the weak anisotropic spin interaction region, an increasing
lattice potential forces a transition from plane-wave to stripe ground
states. The zero momentum pseudo-Goldstone gap first decreases to zero at
the phase transition point and then reopens. In the forced superfluid stripe
region, the properties of gapped excitation spectrum (e.g., zero-momentum
and avoided-crossing gaps) largely resemble those for spontaneous superfluid
stripe phase subject to a weak lattice perturbation (i.e., approximate
symmetry), demonstrating it is a pseudo-Goldstone spectrum. The forced
superfluid stripe ground state is experimentally more accessible and robust
than the spontaneous one, opening the pathway for the direct observation of
pseudo-Goldstone spectrum in experiments.

\textit{iii)} The structure factors of the two lowest energy modes show that
gapless Goldstone and gapped pseudo-Goldstone branches correspond to density
(phonon) and spin-density (magnon) modes, respectively. The static structure
factor for the spin-density reveals some differences between spontaneous and
forced superfluid stripe ground states due to their different stripe
formation mechanisms. The excitation spectrum and structure factor can be
detected in experiments through Bragg scattering.

\begin{figure}[t]
\includegraphics[width=0.48\textwidth]{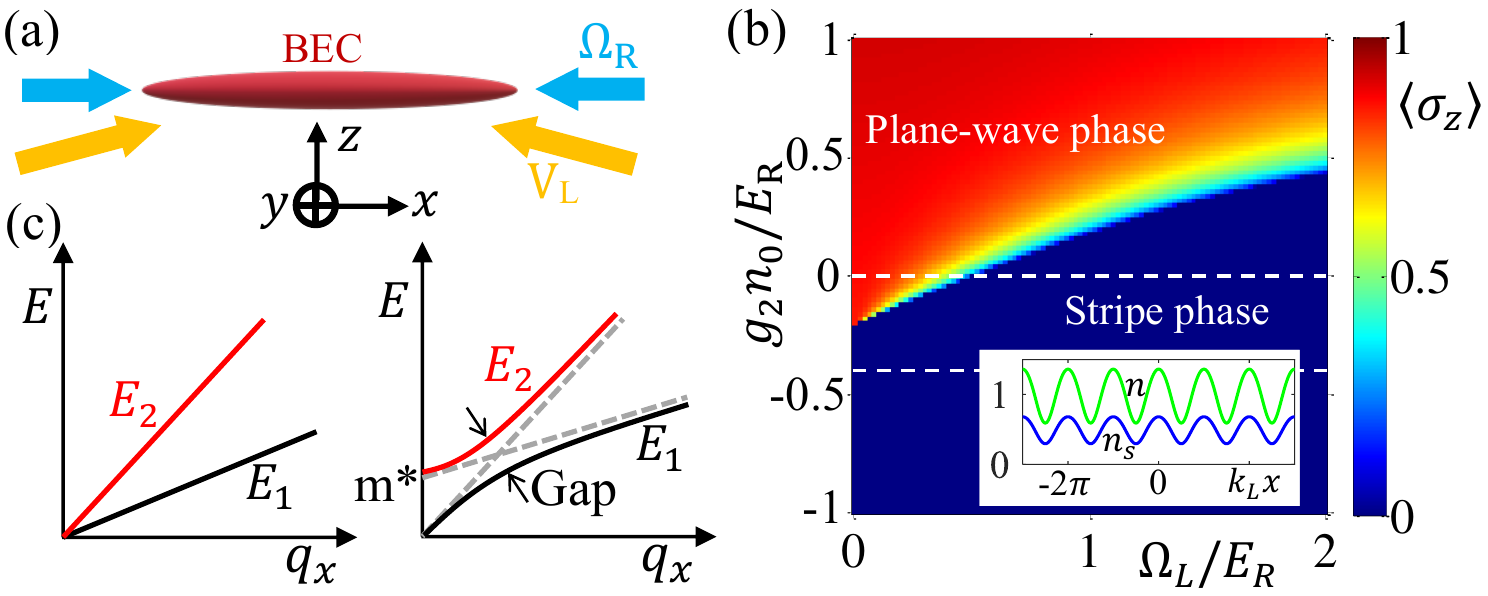}
\caption{(a) Scheme to generate spin-orbit coupling and optical lattice for
a trapped BEC. (b) Phase diagram in the $g_{2}$-$\Omega _{L}$ plane, with $%
\Omega _{R}=2.0E_{R}$, $\protect\delta =0$, $gn_{0}=1.0E_{R}$ and $n_{0}$
the mean atom density. Two bold dashed lines correspond to the weak and
strong spin interactions regimes for Figs.~\protect\ref{figure2}-\protect\ref%
{figure4}. The inset shows the typical densities (normalized to $n_{0}$) of
spin-up (or spin-down) component (blue line) and the total density (green
line). (c) Schematic illustration of the two (pseudo-)Goldstone modes
without (left panel) and with (right panel) a weak and explicitly
symmetry-breaking term.}
\label{figure1}
\end{figure}

{\color{blue}\emph{Model.---}}We consider the experimental setup illustrated
in Fig.~\ref{figure1}a. The BEC is confined in a cigar-shaped optical dipole
trap, with spin-orbit coupling along the $x$ direction realized by two Raman
laser beams, which couple the two pseudospin states $\left\vert \uparrow
\right\rangle $ and $\left\vert \downarrow \right\rangle $ (e.g., $%
|1,-1\rangle $ and $|1,0\rangle $ of $^{87}$Rb atoms within the $F=1$
hyperfine manifold) with momentum kick $2k_{R}$ ($k_{R}$ is the recoil
momentum). In addition, we consider a weak optical lattice $V_{L}(x)=2\Omega
_{L}\sin ^{2}(k_{L}x)$. The single-particle Hamiltonian in the spin basis
(with momentum and energy units as $\hbar k_{R}$ and $\frac{\hbar
^{2}k_{R}^{2}}{2m}$) reads 
\begin{equation}
H_{0}=(i\partial _{x}+\sigma _{z})^{2}-\frac{\delta }{2}\sigma _{z}+\frac{%
\Omega _{R}}{2}\sigma _{x}+{V}_{L}(x),  \label{GPE3D}
\end{equation}%
where $\Omega _{R}$ is the strength of the Raman coupling and $\delta $ is
the detuning of the two-photon Raman transition. Spin-orbit coupling induces
a momentum-space double-well band dispersion, and the period of the optical
lattice is set such that $2k_{L}$ equals to the separation between two band
minima.

We first find the ground state $\psi _{0s}(x)$ by imaginary-time evolution
of the Gross-Pitaevskii (GP) equation 
\begin{equation}
i\frac{\partial \psi _{s}(x,t)}{\partial t}=H_{\text{GP}}(\psi _{s})\psi
_{s}(x,t),  \label{GPE1D}
\end{equation}%
where $\psi _{s}$ is the spinor wavefunction with $s=\uparrow ,\downarrow $
and $H_{\text{GP}}(\psi _{s})=H_{0}+gn+g_{2}|\psi _{\bar{s}}|^{2}$ with $n$
the total density. We have assumed the intra-spin interaction as $%
g_{\uparrow \uparrow }=g_{\downarrow \downarrow }=g$ and the anisotropic
spin interaction as $g_{2}=g_{\uparrow \downarrow }-g$, with $g_{ss^{\prime
}}$ the interaction strength between atoms in $s$ and $s^{\prime }$ states.
The phase diagram in the $g_{2}$-$\Omega _{L}$ plane is shown in Fig.~\ref%
{figure1}b [typical (spin-)density distributions of the stripe state are
shown in the inset]. The system favors the stripe phase (plane-wave phase)
for large negative (positive) $g_{2}$ (note that the stripe and plane-wave
phases become the unpolarized and polarized Bloch states in the presence of
a lattice). The phase boundary lies at the weak anisotropic interaction
region around $g_{2}=0$, and the critical value of $g_{2}$ increases with
the lattice strength. 

Without the optical lattice, the stripe phase occupying both band minima can
be formed in the system under the anti-ferromagnetic atomic interaction ($%
g_{2}<0$) \cite{ChunjiWang2010, TLHo2011, Lin2011, YunLi2012}. Typically,
the stripe phase only exits for very weak $\Omega _{R}$ and $\delta $ due to
the weak anisotropy of interaction $|g_{2}|$ in realistic experiments,
making its observation difficult. This may be overcome by using atoms with
strong anisotropic spin interactions, or alternatively, by adding a weak
optical lattice that couples the two band minima directly. The latter
approach has led to the recent observation of long-lived superstripe state
using $^{87}$Rb atoms~\cite{BersanoExperimental2019}. We want to point out
that, there is a tiny difference between the ground-state stripe period at $%
\Omega _{L}=0$ and the optical lattice period. The two periods would match
as long as the optical lattice strength is not extremely small.

The forced stripe ground state induced by symmetry-breaking potential shows
similar (spin-)density patterns as the spontaneous one induced solely by the
anisotropic interactions. To characterize and distinguish the stripe states
formed under different symmetry breaking mechanisms, we consider the
excitation spectrum. For spontaneous stripe phase induced solely by
interactions, both the $U(1)$ gauge and continuous translational symmetries
are broken spontaneously, leading to two gapless Goldstone modes (as
illustrated in the left panel of Fig.~\ref{figure1}c)~\cite%
{LiSuperstripes2013, XLChen2018, LiChen2017}. When the translational
symmetry is weakly broken by a lattice perturbation (i.e., the symmetry is
now approximate), we expect to observe the gap opening of one Goldstone mode
(equivalent to an effective mass $m^{\ast }$ of the corresponding Goldstone
boson). If the lower mode becomes a gapped pseudo-Goldstone mode, there
should be an avoided crossing due to the hybridization of the two original
Goldstone modes (as illustrated in the right panel of Fig.~\ref{figure1}c).



\begin{figure}[t]
\includegraphics[width=0.48\textwidth]{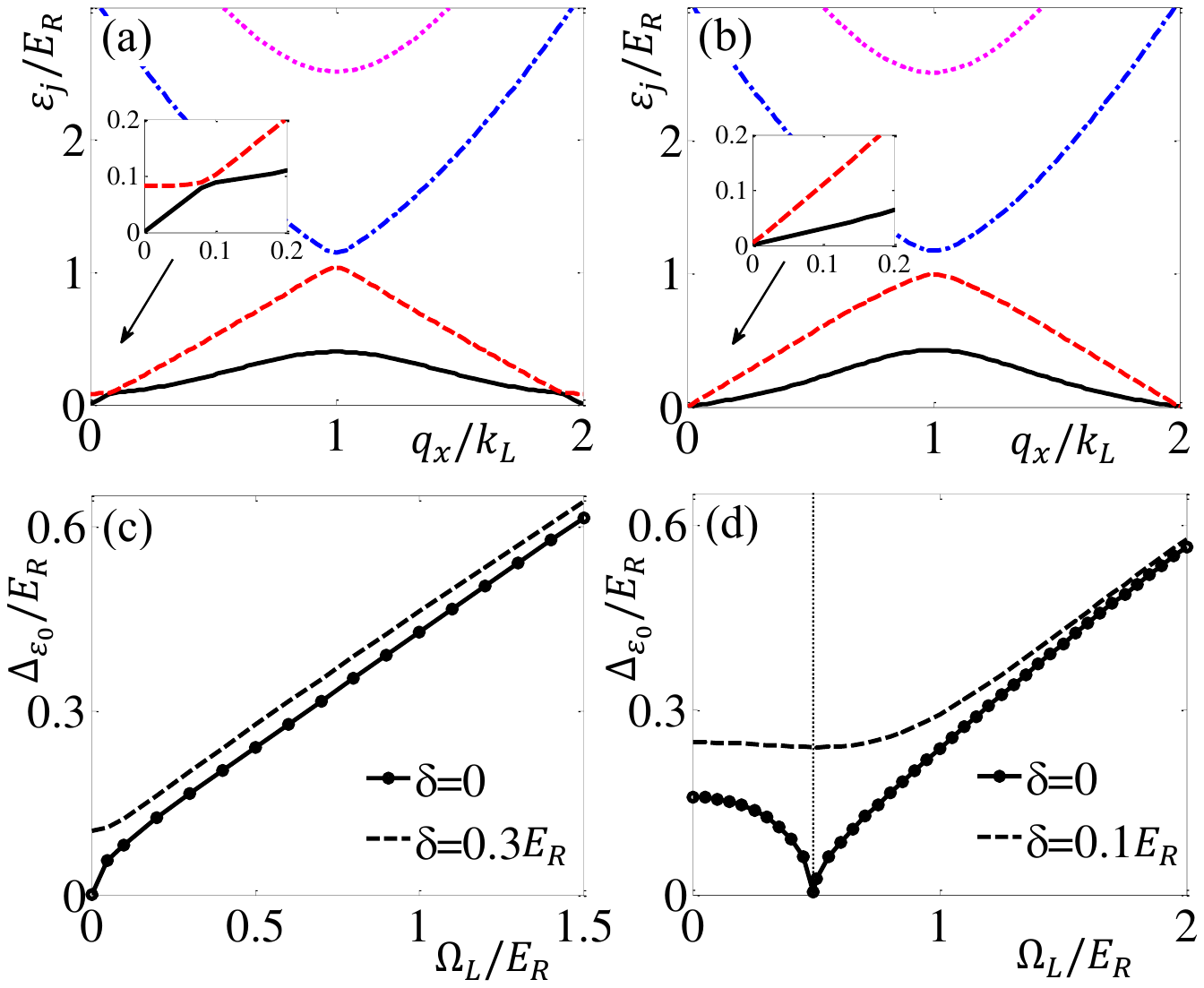}
\caption{(a,b) Low-energy spectra of the elementary excitations for the
stripe phase. (c,d) Change of the zero-momentum gap $\Delta _{\protect%
\varepsilon _{0}}$ with the lattice strength $\Omega _{L}$. (a,c) and (b,d)
are for strong ($g_{2}n_{0}=-0.4E_{R}$) and weak ($g_{2}n_{0}=-0.005E_{R}$)
anisotropic spin interactions, respectively. The lattice strength $\Omega
_{L}=0.1E_{R}$ in (a), and $\Omega _{L}^{c}=0.4871E_{R}$ in (b) is the phase
transition point between the plane-wave phase and the stripe phase. The
dashed lines in (c,d) represent nonzero detunings. The dotted line in (d)
corresponds to $\Omega _{L}^{c}$. $gn_{0}=1.0E_{R}$ and $\Omega
_{R}=2.0E_{R} $.}
\label{figure2}
\end{figure}

To obtain the spectrum of elementary excitations, we write the deviations of
the wavefunctions with respect to the ground state as 
\begin{equation}
\psi _{s}(x,t)=e^{-i\mu t}\left[ \psi _{0s}(x)+u_{s}(x)e^{-i\varepsilon
t}+v_{s}^{\ast }(x)e^{i\varepsilon t}\right] .  \label{Bogoliubov}
\end{equation}%
The amplitudes $u_{s}(x)$ and $v_{s}(x)$ satisfy normalization condition $%
\sum_{s}\int_{0}^{d}dx[|u_{s}(x)|^{2}-|v_{s}(x)|^{2}]=1$, with $d$ the
stripe period and $\mu $ the chemical potential. Substituting Eq.~(\ref%
{Bogoliubov}) into Eq.~(\ref{GPE1D}), we obtain the Bogoliubov equation as $%
\varepsilon \lbrack u_{\uparrow },u_{\downarrow },v_{\uparrow
},v_{\downarrow }]^{T}=\mathcal{H}[u_{\uparrow },u_{\downarrow },v_{\uparrow
},v_{\downarrow }]^{T}$. 
The expression of the Bogoliubov Hamiltonian $\mathcal{H}$ is given in~\cite%
{SM}, and the excitation spectra can be calculated numerically by expanding $%
u_{s}(x)$ and $v_{s}(x)$ in the Bloch basis. Each excitation spectrum is
periodic in momentum space with Brillouin zone determined by the stripe
period.

{\color{blue}\emph{Pseudo-Goldstone spectrum.---}}We focus mainly on the
elementary excitations under the situation of the anti-ferromagnetic atomic
interaction (i.e., $g_{2}<0$), where the stripe phase mainly resides. For a
typical Raman coupling $\Omega _{R}\gtrsim E_{R}$, the system prefers to
form the stripe (plane-wave) phase under strong (weak) anisotropic spin
interaction $|g_{2}|$ in the absence of optical lattices. We first consider
the strong anisotropic spin interaction with a weak optical lattice (lower
region in Fig.~\ref{figure1}b). The optical lattice slightly breaks the
space translational symmetry of the system yet alters the excitation
spectrum dramatically. The low-energy bands in the first Brillouin zone with
weak optical lattice are demonstrated in Fig.~\ref{figure2}a. The double
gapless spectrum disappears and a gap $\Delta _{\varepsilon _{0}}$ in the
second band at zero Bloch momentum ($q_{x}=0$) is opened, which corresponds
to the generation of the pseudo-Goldstone mode of the system at the long
wavelength limit. The pseudo-Goldstone mode is generated once the lattice is
turned on. The change of the zero-momentum gap $\Delta _{\varepsilon _{0}}$
with the strength of the optical lattice is given in Fig.~\ref{figure2}c.
The gap vanishes at zero lattice strength and increases with increasing
optical lattice strength. The size of the gap almost changes linearly with
the optical lattice except near the zero momentum. A small detuning $\delta $
would hardly affect the spectrum, while a large $\delta $ may drive the
system out of the stripe phase and lead to a roton gap ($\Delta
_{\varepsilon _{0}}>0$) at $\Omega _{L}=0$. 
The effect of $\delta $ is diminished at larger lattice strength.

With decreasing anisotropic spin interaction $\left\vert g_{2}\right\vert $,
the system is driven from the stripe phase into the plane-wave phase (i.e.,
the polarized Bloch state). The pseudo-Goldstone gap $\Delta _{\varepsilon
_{0}}$ decreases to zero at the critical phase boundary and then reopens as
a nonzero roton gap. Further increasing lattice strength can drive the
transition from the polarized Bloch state to the stripe phase, where roton
gap $\Delta _{\varepsilon _{0}}$ decreases to zero at the critical phase
boundary and then reopens as the pseudo-Goldstone gap (see Fig.~\ref{figure2}%
d). Notice that at the phase boundary, the excitation spectrum of the forced
stripe state is very similar to that for strong anisotropic interaction with
two gapless Goldstone modes (see Fig.~\ref{figure2}b). For weak anisotropic
spin interactions ($g_{2}\simeq 0$), the phase transition boundary locates
at the lattice strength $\Omega _{L}^{c}=0.4871E_{R}$. Beyond the critical
lattice strength $\Omega _{L}^{c}$, the pseudo-Goldstone gap (see Fig.~\ref%
{figure2}d) behaves similarly as that in the strong anisotropic interaction
regime (see Fig.~\ref{figure2}c), demonstrating the properties of the
pseudo-Goldstone spectrum. Therefore, the forced superfluid stripe ground
state, which is experimentally more accessible and robust than the
spontaneous one, opens the pathway for the direct observation of
pseudo-Goldstone spectrum in experiments utilizing techniques that have
already been used to study the spectrum of elementary excitations for
spin-orbit-coupled BECs~\cite{PhysRevA.90.063624,PhysRevLett.114.105301},
quantum gas with cavity-mediated long-range interactions~\cite{R.Mottl2012}
and BEC in a shaken optical lattice~\cite{PhysRevLett.114.055301}.

The pseudo-Goldstone mode near the phase boundary results from the interplay
between the interaction and optical lattice, which tends to reduce the
spatial modulation of the GP Hamiltonian $H_{\text{GP}}(\psi _{0s})$ for the
ground state. In the vicinity of the phase boundary, $H_{\text{GP}}(\psi
_{0s})$ preserves an approximate translational symmetry \cite{SM}, which
leads to a vanishing gap of the pseudo-Goldstone mode. On the other hand,
the pseudo-Goldstone mode for strong $|g_{2}|$ with a very weak lattice is
induced by the approximate translational symmetry of $H_{0}$, while $H_{%
\text{GP}}(\psi _{0s})$ strongly breaks the translational symmetry. In the
presence of nonzero detuning $\delta $, the phase boundary becomes a
crossover boundary, and $\Delta _{\varepsilon _{0}}$ decreases to a finite
value as the system goes from the stripe region to the plane-wave region,
where $\Delta _{\varepsilon _{0}}$ is almost a constant (see Fig.~\ref%
{figure2}d). The effects of the Raman coupling and the atomic interactions
on $\Delta _{\varepsilon _{0}}$ are given in~\cite{SM}.

\begin{figure}[t]
\includegraphics[width=0.47\textwidth]{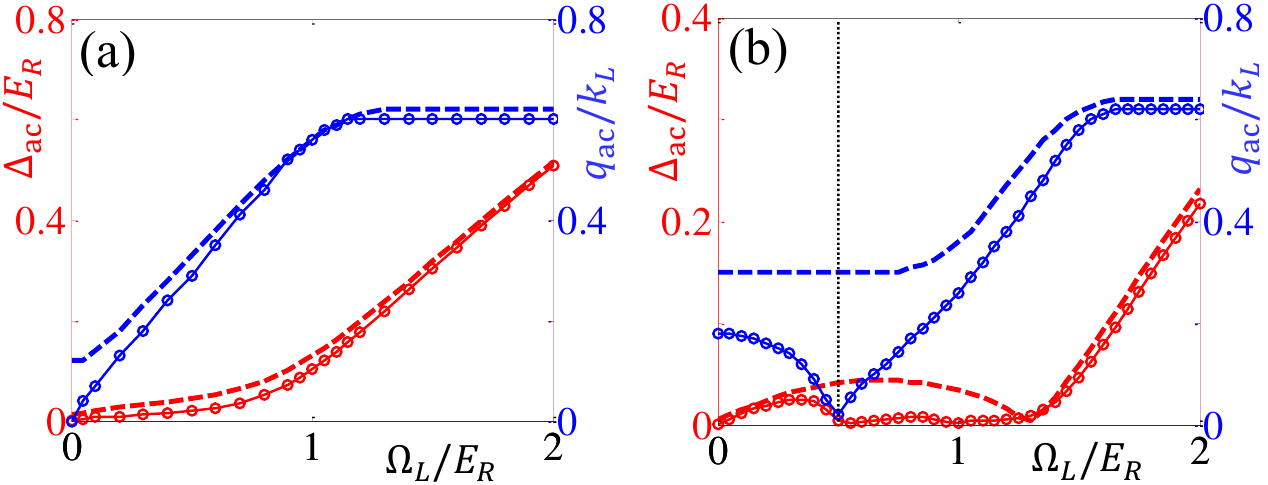}
\caption{Change of the size $\Delta _{\text{ac}}$ and position $q_{\text{ac}%
} $ of the nonzero-momentum gap with the lattice strength $\Omega _{L}$. The
parameters in (a) and (b) are the same as those in Figs.~\protect\ref%
{figure2}(c) and \protect\ref{figure2}(d). The dashed lines represent
nonzero detuning $|\protect\delta |=0.3E_{R}$ in (a) and $|\protect\delta %
|=0.1E_{R}$ in (b), respectively. The gray dotted line in (b) corresponds to
the phase transition point $\Omega _{L}^{c}=0.4871E_{R}$.}
\label{figure3}
\end{figure}

{\color{blue}\emph{Avoided spectrum crossing.---}}In addition to the
zero-momentum gap, there exists another avoided crossing gap $\Delta _{\text{%
ac}}$ (the minimum value of the gap between the first and second bands),
originating from the hybridization between the pseudo-Goldstone and
Goldstone modes. The nonzero-momentum gap $\Delta _{\text{ac}}$ as a
function of the lattice strength for strong anisotropic spin interaction is
given in Fig.~\ref{figure3}a. The gap increases slowly with increasing
lattice strength at the beginning, and then rapidly in the deep lattice
region. In contrast, the avoided crossing point $q_{\text{ac}}$ first
increases rapidly with the lattice strength, and remains saturated in the
deep lattice region. Fig.~\ref{figure3}b shows $\Delta _{\text{ac}}$ as a
function of lattice strength for weak anisotropic spin interaction. In the
plane-wave phase, $\Delta _{\text{ac}}$ first increases with the lattice
strength and then decreases to zero at the phase boundary, while $q_{\text{ac%
}}$ decreases directly to zero. In the stripe phase, $\Delta _{\text{ac}}\ $
and $q_{\text{ac}}$ behave similarly as those for the strong anisotropic
spin interaction (see Fig.~\ref{figure3}a). A nonzero Raman detuning $\delta 
$ tends to increase both the gap $\Delta _{\text{ac}}$ and its position $q_{%
\text{ac}}$ (see the dashed lines in Fig.~\ref{figure3}a and \ref{figure3}%
b). The effect of the Raman coupling on $\Delta _{\text{ac}}\ $ and $q_{%
\text{ac}}$ are given in~\cite{SM}.

\begin{figure}[t]
\includegraphics[width=0.48\textwidth]{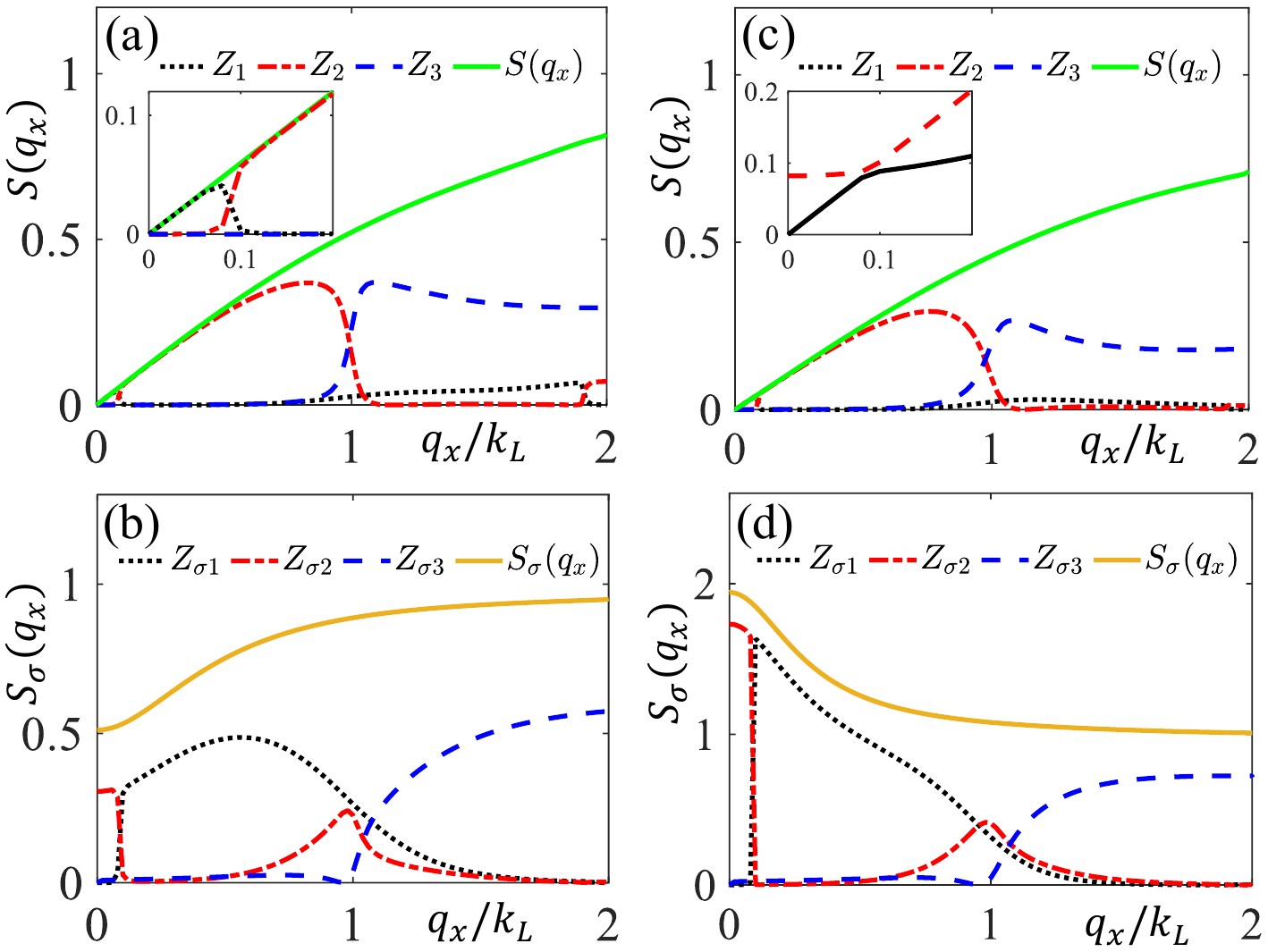}
\caption{Static structure factors and excitation strengths for the density
in (a,c) and spin-density in (b,d). (a,b) correspond to the excitation
spectrum of the stripe phase in Fig.~\protect\ref{figure2}(a), and (c,d)
correspond to the spectrum under the same parameter condition as Fig.~%
\protect\ref{figure2}(b) except $\Omega _{L}=0.6E_{R}$. The insets in (a)
and (c) show the corresponding excitation strengths and spectra near $q_x=0$%
, respectively. }
\label{figure4}
\end{figure}

{\color{blue}\emph{Structure factors.---}}As discussed above, the
pseudo-Goldstone spectra for the weak and strong anisotropic spin
interactions are very similar, although the ground stripe phases are
achieved through different symmetry breaking mechanisms. In experiments, the
collective properties of the excitation spectrum of the BECs can be probed
using Bragg spectroscopy, which measures the dynamical structure factors.
For a scattering probe with momentum $\hbar q_x$ and energy $\hbar \omega $,
the dynamical structure factor takes the form~\cite{LiSuperstripes2013,
LP.SS.Book2016}: 
\begin{equation}
S(q_x,\omega )=\sum\nolimits_{j}|\langle j|\rho _{q_x}^{\dagger }|0\rangle
|^{2}\tilde{\delta}(\hbar \omega -\varepsilon _{j})  \label{DSM}
\end{equation}%
with $|j\rangle $ the excited state, $\varepsilon _{j}$ the excitation
energy, $\rho _{q_x}=\sum_{j}e^{iq_xx_{j}/\hbar }$ the density operator and $%
\tilde{\delta}(\cdot )$ the Dirac delta function. The excitation strength $%
Z_{j}=|\langle j|\rho _{q_x}^{\dagger }|0\rangle |^{2}$ can be evaluated as: 
\begin{equation}
Z_{j}=\sum_{s}\left\vert \int_{0}^{d}[u_{js}^{\ast }(x)+v_{js}^{\ast
}(x)]e^{iq_{x}x}\psi _{0}(x)dx\right\vert ^{2}.  \label{ZM}
\end{equation}%
The integral of the dynamical structure factor gives the static density
structure factor $S(q_x)=\int S(q_x,\omega )d\omega $, which is uniquely
determined by the sum of the excitation strengths for all energy bands.
Similarly, we can define the spin-density static structure factor $S_{\sigma
}(q_{x})$ and the excitation strength $Z_{\sigma j}$ by replacing $\rho
_{q_x}^{\dagger }$ with $\sigma _{z}\rho _{q_x}^{\dagger } $. Such spin
structure factors may be probed by spin-dependent Bragg spectroscopy using
lasers with suitable polarization and detuning~\cite%
{Spin.Bragg1,Spin.Bragg2,Spin.Bragg3}. $S(q_{x})$ and $S_{\sigma }(q_{x})$
are related to density and spin-wave excitations, respectively. Both of them
include the contributions from all of considered energy bands.

The static structure factors for strong anisotropic spin interaction with a
weak lattice $\Omega _{L}=0.1E_{R}$ are given in Fig.~\ref{figure4}a for the
density and Fig.~\ref{figure4}b for the spin-density, where the excitation
strengths $Z_{j}$ and $Z_{\sigma j}$ for the first three excitation bands
are also shown. $S(q_x)$ and $S_{\sigma }(q_{x})$ increase with the
quasi-momentum $q_{x} $ monotonically. $S(q_x)$ vanishes, but $S_{\sigma
}(q_{x})$ has a non-zero minimum value at $q_{x}=0$. The excitation
strengths for the first and second bands exchange at the position of the
nonzero-momentum gap $q_\text{ac}=0.09k_{L}$, showing that the first and
second lowest bands correspond to density and spin excitations,
respectively. This feature could be used to identify the pseudo-Goldstone
modes in Bragg spectroscopy experiments.

$S(q_{x})$ in the forced stripe phase for the weak anisotropic spin
interaction (see Fig.~\ref{figure4}c) has similar features as Fig.~\ref%
{figure4}a, while $S_{\sigma }(q_{x})$ (see Fig.~\ref{figure4}d) shows quite
different properties from Fig.~\ref{figure4}b. In the forced stripe phase, $%
S_{\sigma }(q_{x})$ decreases monotonically with $q_{x}$ with a maximum at $%
q_{x}=0$ (see Fig.~\ref{figure4}d). At the phase transition point, $%
S_{\sigma }(q_{x})$ diverges as $q_{x}\rightarrow 0$. The dependence of $%
S(q_{x})$ and $S_{\sigma }(q_{x})$ on other parameters are shown in~\cite{SM}%
.

{\color{blue}\emph{Conclusion.---}}In summary, we show that the collective
excitation spectrum of a spin-orbit-coupled BEC can be used to distinguish
the spontaneous and forced stripe ground states induced by different
symmetry breaking mechanisms. The lattice forced stripe phase, which is
experimentally more accessible and robust than the spontaneous one, can
provide direct experimental evidence for the long-sought gapped
pseudo-Goldstone spectrum. While the present work focuses on
spin-orbit-coupled BECs, similar ideas can be implemented to other systems
such as dipolar striped BECs~\cite{PhysRevLett.122.130405,
PhysRevX.9.011051,PhysRevX.9.021012,NataleExcitation2019}, striped BECs in
optical superlattice~\cite{LiA2017} or in a cavity~\cite{cinti2014defect,
baumann2010dicke, PhysRevLett.113.070404, leonard2017supersolid}, to study
pseudo-Goldstone modes in the presence of approximate/forced symmetry
breaking.

\begin{acknowledgments}
\textbf{Acknowledgements}: We thank P. Engels, S. Mossman, E. Crowell, and
T. Bersano for helpful discussions. X.W.L., J.H. and C.Z. are supported by
Air Force Office of Scientific Research (FA9550-20-1-0220), National Science
Foundation (PHY-1806227), and Army Research Office (W911NF-17-1-0128).
G.Q.L. acknowledges the supports from NSF of China (Grant No.~11405100), the
Natural Science Basic Research Plan in Shaanxi Province of China (Grant
Nos.~2019JM-332 and 2020JM-507), the Doctoral Research Fund of Shaanxi
University of Science and Technology in China (Grant No.~2018BJ-02), and the
China Scholarship Council (Program No.~201818610099).
\end{acknowledgments}

\onecolumngrid

\newpage \clearpage
\onecolumngrid
\appendix

\section*{Supplementary Material for ``Pseudo-Goldstone Excitations in a
Striped Bose-Einstein Condensate"}

\setcounter{table}{0} \renewcommand{\thetable}{S\arabic{table}} %
\setcounter{figure}{0} \renewcommand{\thefigure}{S\arabic{figure}} %
\setcounter{equation}{0} \renewcommand{\theequation}{S\arabic{equation}}

\bigskip

\bigskip {\color{blue}\emph{Method.---}} In order to calculate the
excitation spectrum of the spontaneous and forced stripe phases, we first
find the ground states of the system. We adopt the following ansatz 
\begin{equation}
\left( 
\begin{array}{c}
\psi _{0\uparrow }(x) \\ 
\psi _{0\downarrow }(x)%
\end{array}%
\right) =\sum_{K}\left( 
\begin{array}{c}
a_{K} \\ 
-b_{K}%
\end{array}%
\right) e^{i(K+k_{0})x},
\end{equation}%
where $K=(2l-1)k_{L}$ with the integer $l=-L,...,L+1$ represent the
reciprocal lattice vectors, $L$ is the cutoff of the plane-wave modes. The
expansion coefficients $a_{K}$ and $b_{K}$, together with $k_{0}$, are
determined by minimizing the energy functional 
\begin{equation}
E=\int dx\psi ^{\dagger }(x)[H_{0}+\frac{1}{2}H_{\text{int}}(\psi
_{0\uparrow },\psi _{0\downarrow })]\psi (x),
\end{equation}%
where $\psi =(\psi _{0\uparrow },\psi _{0\downarrow })^{T}$ is the
two-component spinor wavefunction normalized by the atom number $N=\int
dx\psi ^{\dagger }(x)\psi (x)$, and $H_{\text{int}}(\psi _{0\uparrow },\psi
_{0\downarrow })=\mathrm{diag}[gn+g_{2}|\psi _{0\downarrow }|^{2},g_{2}|\psi
_{0\uparrow }|^{2}+gn]$ with density $n=|\psi _{0\uparrow }|^{2}+|\psi
_{0\downarrow }|^{2}$. The results for the ground state are calculated
numerically by the imaginary-time evolution of the Gross-Pitaevskii (GP)
equation. The corresponding initial solution is given by the variational
method considering the lowest-four modes in the wavefunction ansatz.

To evaluate the spectrum of elementary excitations, the Bogoliubov equation
is obtained by writing the deviation of the wavefunction with respect to the
ground state as 
\begin{equation}
\psi _{s}(x,t)=e^{-i\mu t}\left[ \psi _{0s}(x)+u_{s}(x)e^{-i\varepsilon
t}+v_{s}^{\ast }(x)e^{i\varepsilon t}\right] .  \label{Bogoliubov0}
\end{equation}%
The perturbation amplitudes $u_{s}(x)$ and $v_{s}(x)$ with $s=\uparrow
,\downarrow $ are expanded in the Bloch form in terms of the reciprocal
lattice vectors: 
\begin{equation}
u_{s}(x)=\sum_{m=-M}^{M+1}U_{s,m}e^{i(k_0+q_x)x+i(2m-1)k_{L}x},
\label{Bogoliubov1}
\end{equation}%
\begin{equation}
v_{s}(x)=\sum_{m=-M}^{M+1}V_{s,m}e^{i(k_0+q_x)x-i(2m-1)k_{L}x},
\label{Bogoliubov2}
\end{equation}%
where $q_x$ is the Bloch wavevector of the excitations and $M$ is the cutoff
of the plane waves of the excited states.


{\color{blue}\emph{Ground state and phase diagram.---}} Depending on the
spin-orbit coupling and the atomic interactions, the spin-orbit-coupled BEC
without the optical lattice potential has three different phases: stripe,
plane-wave and zero-momentum phases~\cite{ChunjiWang2010,
TLHo2011,Lin2011,YunLi2012,LiSuperstripes2013}. 
The parameter range for the existence of the stripe phase is very narrow,
following with the small contrast and small wavelength of the fringes, which
make the observation of the stripe state very difficult in experiments. In
contrast, the stripe phase in the spin-orbit-coupled BEC forced by the weak
optical lattice has been observed recently~\cite{BersanoExperimental2019}.
The key feature is that the wavelength of the lattice beams and the Raman
coupling strength are chosen such that the lattice couples two minima of the
lower spin-orbit band, where the static spin-independent lattice provides a $%
2k_{L}$ momentum kick while preserves the spin. 
Such forced stripe state has a long lifetime and is more stable, and its
existing parameter region is extended dramatically.

\begin{figure}[t]
\includegraphics[width=0.7\textwidth]{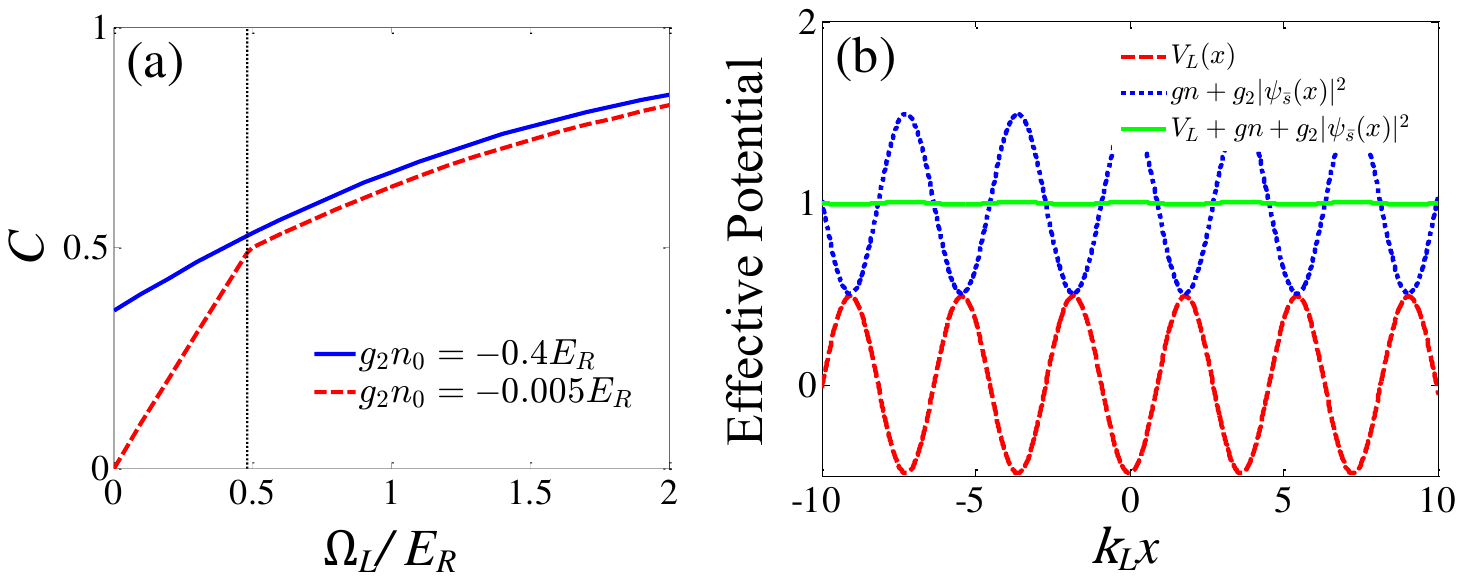}
\caption{(a) The modulation contrast of the ground state as a function of
the strength of the optical lattice. The vertical dotted line represents the
phase boundary between the plane-wave phase and the stripe phase for $%
g_{2}n_{0}=-0.005E_{R}$. (b) The effective potential (i.e., $gn+g_2|\protect%
\psi_{\bar{s}}|^2+V_L$) of $H_\text{GP}(\protect\psi_{0s})$ at the phase
boundary. Other parameters are $\Omega _{R}=2.0E_{R}$, $\protect\delta =0$
and $gn_{0}=1.0E_{R}$.}
\label{FigureS1}
\end{figure}

With a large Raman coupling strength like $\Omega _{R}=2.0E_{R}$ and without
the optical lattice, the stripe phase and plane wave phase appear at strong
and weak anisotropic spin interaction regions, respectively. With the
optical lattice, there exists a magnetized feature related with the
plane-wave phase (i.e., the polarized Bloch state) in the system, which was
also revealed in previous studies~\cite{GIMartone2016, ZChen2016}. The
stripe phase (i.e., unpolarized Bloch wave) for the two components exists at
larger optical lattice strengths. In the formation of the stripe phase, the
modulation depth of the density increases with the increasing lattice
strength. The contrast of the total density $%
C=(n_{max}-n_{min})/(n_{max}+n_{min})$ reflects this change and shows the
phase transition between the polarized Bloch state and the unpolarized Bloch
state (the perfect stripe phase) for the weak anisotropic spin interaction
[Fig.~\ref{FigureS1}(a)].

In Fig.~1(b) of the main text, the spin polarization is plotted with respect
to the optical lattice strength and the anisotropic spin interaction between
atoms. For the strong anisotropic spin interaction, the existence of the
stripe phase does not need the optical lattice. For the weak anisotropic
spin interaction, there is a critical lattice strength $\Omega _{L}^{c}$ ($%
\Omega _{L}^{c}=0.4871E_{R}$ for $gn_{0}=1.0E_{R}$ and $%
g_{2}n_{0}=-0.005E_{R}$) for the transition from the plane-wave phase (i.e.,
polarized Bloch state with $\langle \sigma _{z}\rangle \neq 0$) to the
stripe phase (i.e., the unpolarized Bloch state with $\langle \sigma
_{z}\rangle =0$). We calculate the first- and second-order derivatives of
the ground-state energy $E$ with respect to the lattice strength $\Omega
_{L} $. The jump in the second order derivative with $\Omega _{L}$ shows
that the phase transition is the second-order. The phase transition point
can also be identified from the excitation spectrum as discussed in the main
text.

At the phase transition point $\Omega _{L}^{c}$, the spatial modulation due
to the atom density in the GP Hamiltonian $H_{\text{GP}}(\psi _{0s})$
cancels with the spatial modulation of the lattice potential $V_{L}\left(
x\right) $ [see Fig.~\ref{FigureS1}(b)], therefore $H_{\text{GP}}(\psi _{0s})
$ is close to a constant at the ground state. $H_{\text{GP}}(\psi _{0s})$
preserves the translational symmetry, leading to two gapless Goldstone modes
shown in the inset of Fig. 2(b) in the main text. Slightly above $\Omega
_{L}^{c}$, the translational symmetry becomes approximate, leading to the
pseudo-Goldstone gap. In this context, the region above $\Omega _{L}^{c}$
for the forced stripe phase resembles the spontaneous stripe phase with a
very weak lattice perturbation (\textit{i.e.}, approximate symmetry region). 

{\color{blue}\emph{Bogoliubov equations.---}} By substituting Eqs.~(\ref%
{Bogoliubov0})-(\ref{Bogoliubov2}) into Eq.~(2) in the main text, the
Bogoliubov equation can be obtained as follows: 
\begin{equation}
\mathcal{H}\left( 
\begin{array}{c}
u_{\uparrow } \\ 
u_{\downarrow } \\ 
v_{\uparrow } \\ 
v_{\downarrow }%
\end{array}%
\right) =\varepsilon \left( 
\begin{array}{c}
u_{\uparrow } \\ 
u_{\downarrow } \\ 
v_{\uparrow } \\ 
v_{\downarrow }%
\end{array}%
\right) ,  \label{BdGE}
\end{equation}%
where the Bogoliubov Hamiltonian 
\begin{equation}
\mathcal{H}=\left( 
\begin{array}{cccc}
\mathcal{H}_{\uparrow } & \frac{\Omega _{R}}{2}+g_{\uparrow \downarrow }\psi
_{0\uparrow }\psi _{0\downarrow }^{\ast } & g\psi _{0\uparrow }^{2} & 
g_{\uparrow \downarrow }\psi _{0\uparrow }\psi _{0\downarrow } \\ 
\frac{\Omega _{R}}{2}+g_{\uparrow \downarrow }\psi _{0\uparrow }^{\ast }\psi
_{0\downarrow } & \mathcal{H}_{\downarrow } & g_{\uparrow \downarrow }\psi
_{0\uparrow }\psi _{0\downarrow } & g\psi _{0\downarrow }^{2} \\ 
-g\psi _{0\uparrow }^{\ast 2} & -g_{\uparrow \downarrow }\psi _{0\uparrow
}^{\ast }\psi _{0\downarrow }^{\ast } & -\mathcal{H}_{\uparrow }^{\ast } & -%
\frac{\Omega _{R}}{2}-g_{\uparrow \downarrow }\psi _{0\uparrow }^{\ast }\psi
_{0\downarrow } \\ 
-g_{\uparrow \downarrow }\psi _{0\uparrow }^{\ast }\psi _{0\downarrow
}^{\ast } & -g\psi _{0\downarrow }^{\ast 2} & -\frac{\Omega _{R}}{2}%
-g_{\uparrow \downarrow }\psi _{0\uparrow }\psi _{0\downarrow }^{\ast } & -%
\mathcal{H}_{\downarrow }^{\ast }%
\end{array}%
\right) ,  \label{BdGE2}
\end{equation}%
with 
\begin{equation}
\mathcal{H}_{\uparrow }=-\partial ^{2}/\partial x^{2}+2i\partial /\partial
x-\delta /2+V_{L}(x)-\mu +2g|\psi _{0\uparrow }|^{2}+g_{\uparrow \downarrow
}|\psi _{0\downarrow }|^{2},  \label{Hup}
\end{equation}%
\begin{equation}
\mathcal{H}_{\downarrow }=-\partial ^{2}/\partial x^{2}-2i\partial /\partial
x+\delta /2+V_{L}(x)-\mu +2g|\psi _{0\downarrow }|^{2}+g_{\uparrow
\downarrow }|\psi _{0\uparrow }|^{2},  \label{Hdown}
\end{equation}%
and $g_{\uparrow \downarrow }=g+g_{2}$. The time-independent GP equation
becomes%
\begin{equation}
\mu \psi _{0}=\left[ H_{0}+H_{\text{int}}(\psi _{0\uparrow },\psi
_{0\downarrow })\right] \psi _{0}.  \label{TIGPE}
\end{equation}%
The ground state $\psi _{0}$ and the chemical potential $\mu $ are obtained
by the imaginary-time evolution. The Bogoliubov excitation energy $%
\varepsilon $ with respect to $q_{x}$ is numerically obtained by
diagonalizing the Bogoliubov Hamiltonian. 
\begin{figure}[t]
\includegraphics[width=0.7\textwidth]{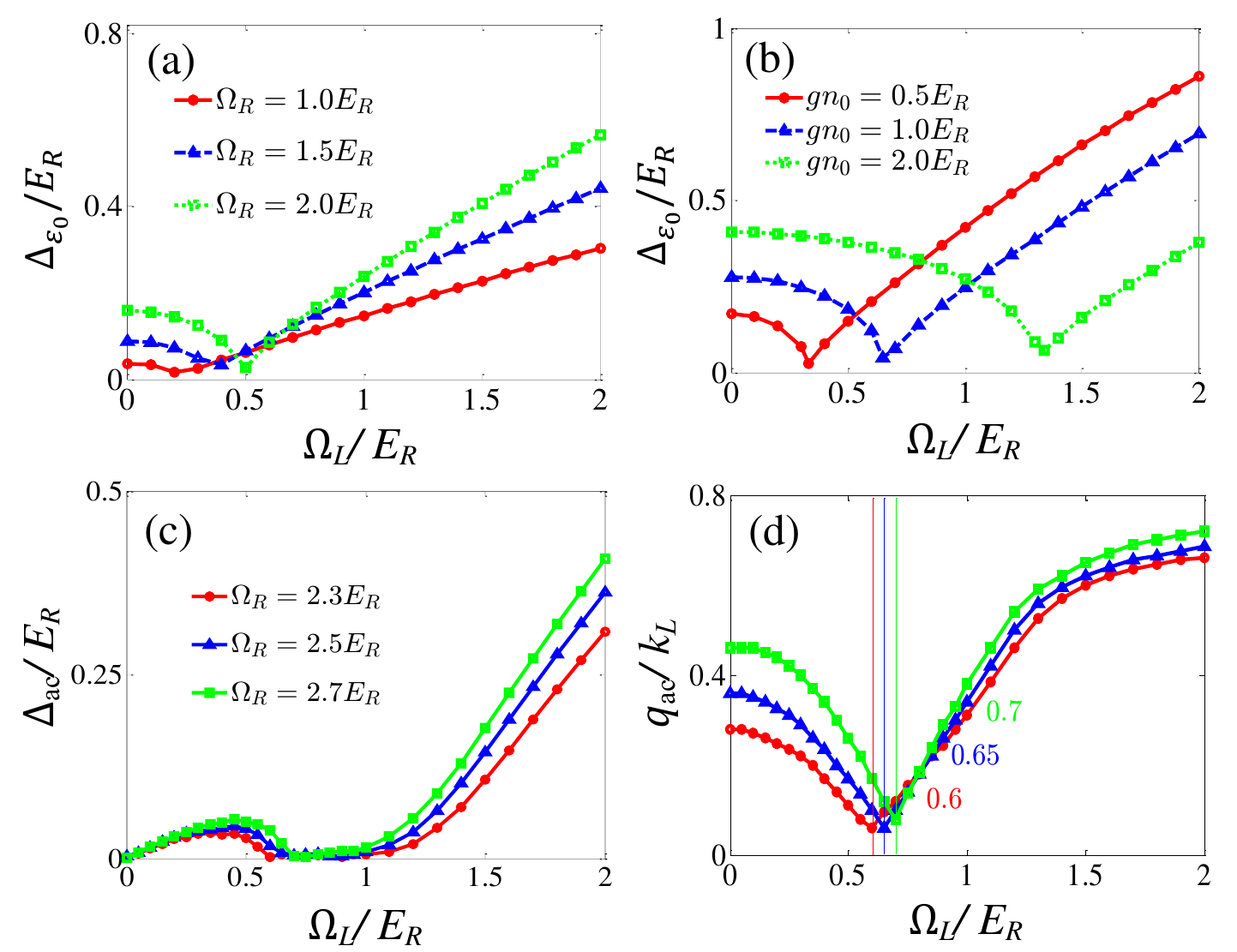}
\caption{(a,b) The zero-momentum gap $\Delta _{\protect\varepsilon _{0}}$ as
a function of lattice strength $\Omega _{L}$ for different $\Omega _{R}$ (a)
and $gn_{0}$ (b). (c,d) $\Delta _{\text{ac}} $ (c) and $q_{\text{ac}}$ (d)
of the avoided crossing gap as functions of $\Omega _{L}$ for $\Omega
_{R}=2.3E_{R}$ (dots), $\Omega _{R}=2.5E_{R}$ (triangles) and $\Omega
_{R}=2.7E_{R}$ (squares), $q_{\text{ac}}$ reaches it minimum at $\Omega
_{L}=0.6E_{R}$, $0.65E_{R}$ and $0.7E_{R} $, respectively. The other
parameters are $\protect\delta =0$, $gn_{0}=1.0E_{R}$ and $g_{2}=-0.005E_{R}$%
.}
\label{FigureS2}
\end{figure}

\begin{figure}[b]
\includegraphics[width=0.6\textwidth]{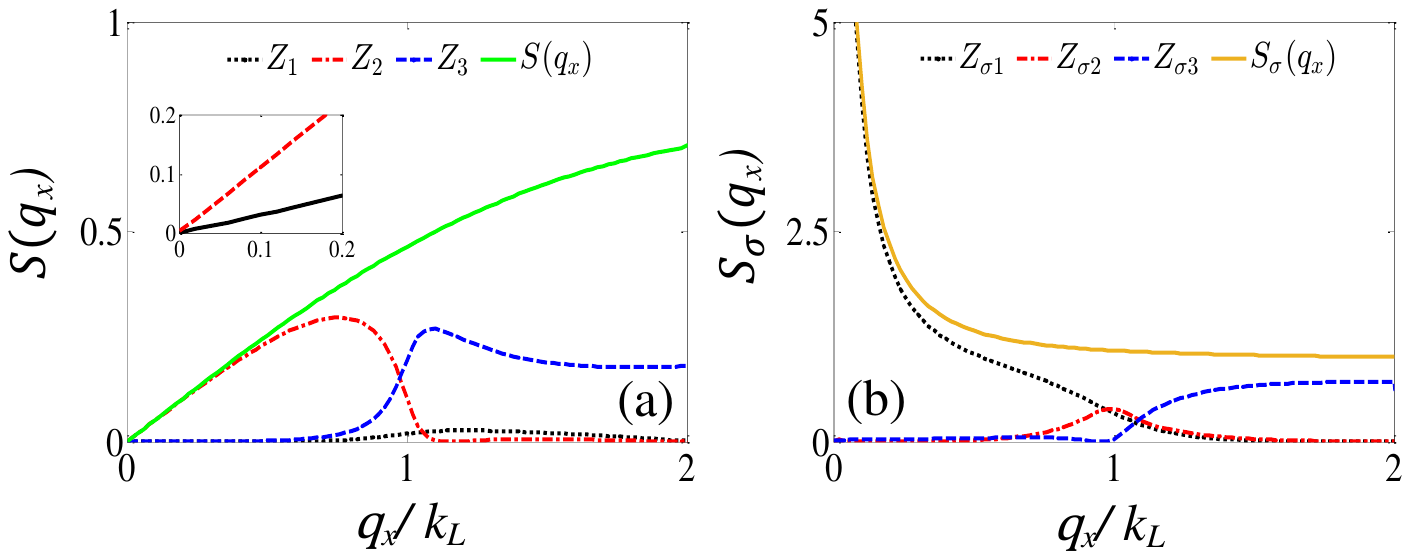}
\caption{Static structure factors and excitation strengths for density (a)
and spin-density (b) at the critical lattice strength $\Omega
_{L}^{c}=0.4871E_{R}$. The insert in (a) corresponds to two lowest energy
bands in the small momentum region. Other parameters are $\Omega
_{R}=2.0E_{R}$, $gn_{0}=1.0E_{R}$ and $g_{2}n_{0}=-0.005E_{R}$.}
\label{FigureS3}
\end{figure}

Besides the Raman detuning $\delta $ demonstrated in Figs.~2c, 2d and Fig.~3
in the main text, the effects of other tunable parameters (Raman coupling
and atomic interactions) on the elementary excitations are shown in Fig.~\ref%
{FigureS2}. The effect of the Raman coupling on zero-momentum gap $\Delta
_{\varepsilon _{0}}$ is shown in Fig.~\ref{FigureS2}(a), where the critical
lattice strength for the minimum of the zero-momentum gap increases with the
increasing Raman coupling strength. The effect of the atomic interactions on 
$\Delta _{\varepsilon _{0}}$ is shown in Fig.~\ref{FigureS2}(b), where the
curves are shifted to larger optical lattice strength when the atomic
interaction $gn_{0}$ increases. The effects of the Raman coupling on the
size and position of nonzero-momentum gap $\Delta _\text{ac}$ are shown in
Figs.~\ref{FigureS2}(c) and \ref{FigureS2}(d). The size of $\Delta _\text{ac}
$ increases with the increasing Raman coupling strength except at some
crossing points. The minimum position of the nonzero-momentum gap shifts to
the larger optical lattice strength for larger Raman coupling strength.

\begin{figure}[t]
\includegraphics[width=0.75\textwidth]{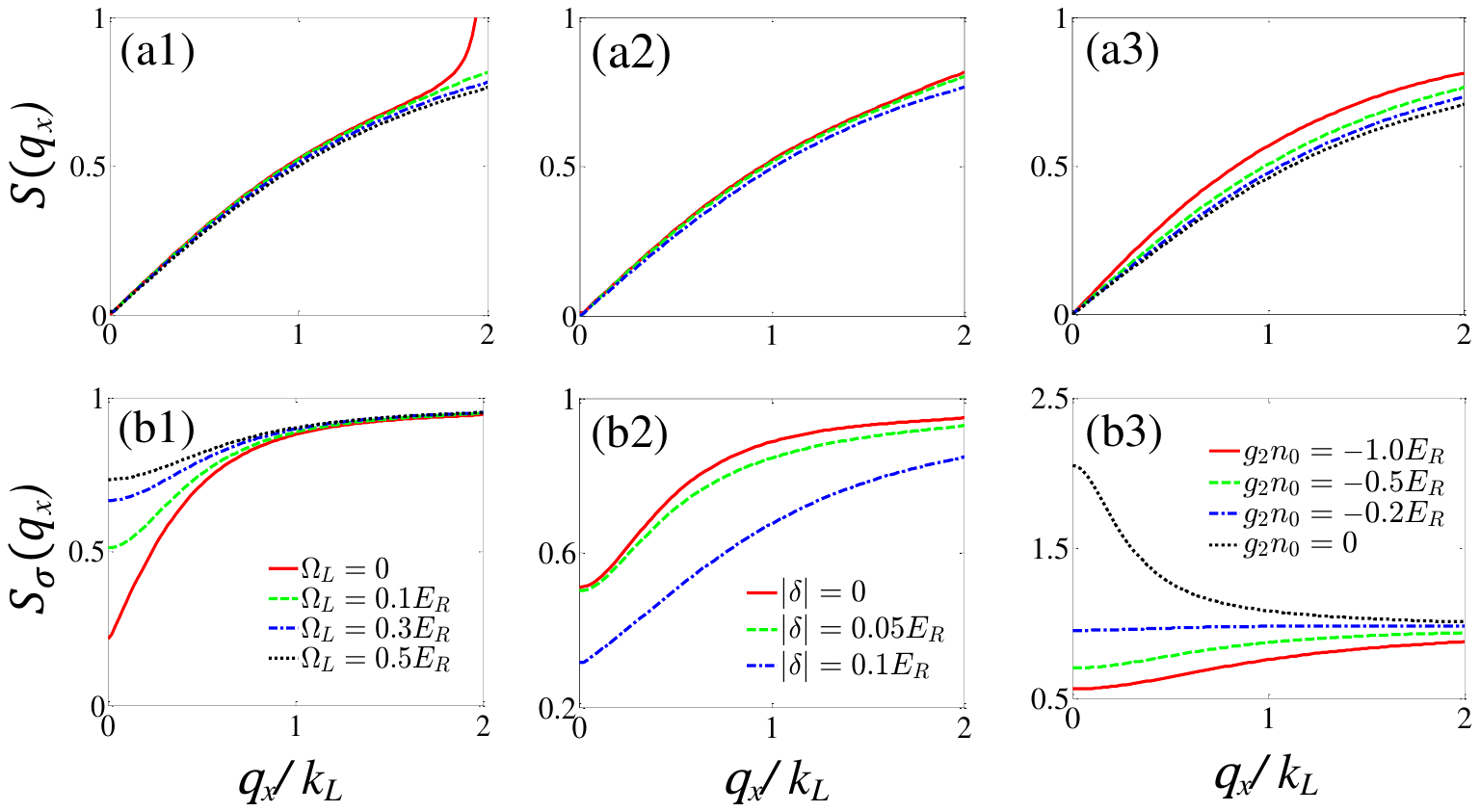}
\caption{Change of the static structure factors for density (a1-a3) and
spin-density (b1-b3) with different parameters $\Omega _{L}$, $\protect%
\delta $ and $g_{2}n_{0}$. (a1, b1) $\protect\delta =0 $, $\Omega
_{R}=2.0E_{R}$, $gn_{0}=1.0E_{R}$ and $g_{2}n_{0}=-0.4E_{R}$; (a2, b2) $%
\Omega _{R}=2.0E_{R}$, $\Omega _{L}=0.1E_{R}$, $gn_{0}=1.0E_{R}$ and $%
g_{2}n_{0}=-0.4E_{R}$; (a3, b3) $\protect\delta =0$, $\Omega _{R}=2.0E_{R}$, 
$\Omega _{L}=0.6E_{R}$ and $gn_{0}=1.0E_{R}$.}
\label{FigureS4}
\end{figure}
{\color{blue}\emph{Dynamical structure factor and Bragg spectroscopy
measurement.---}} The Bragg spectroscopy measures the dynamical structure
factor of the BEC, \textit{i.e.}, the density response of the system to the
external perturbation generated by the scattering probe of momentum $\hbar
q_x$, and energy $\hbar \omega $~\cite{LiSuperstripes2013, LP.SS.Book2016}.
Denoting the linear perturbation $V_{1}=\frac{V}{2}[\rho _{q_x}^{\dagger
}e^{-i\omega t}+\rho _{-q_x}e^{i\omega t}]$, where $\rho
_{q_x}=\sum_{j}e^{iq_xx_{j}/\hbar }$ is the Fourier transformation of
one-body density operator with the probe momenta $q_x$, the dynamical
structure factor takes the form: 
\begin{equation}
S(q_x,\omega )=\sum_{j}|\langle j|\rho _{q_x}^{\dagger }|0\rangle |^{2}%
\tilde{\delta}[\hbar \omega -(E_{j}-E_{0})].  \label{DSSF}
\end{equation}%
Here, $|0\rangle $ ($|j\rangle $) is the ground (excited) state with the
energy $E_{0}$ ($E_{j}$). We can define the spin structure factor $S_{\sigma
}$ in a similar way, which can be measured using spin-dependent Bragg
spectroscopy~\cite{Spin.Bragg1,Spin.Bragg2,Spin.Bragg3}. 

The static structure factors and excitation strengths for the density and
spin-density are given in Fig.~\ref{FigureS3}(a) and \ref{FigureS3}(b) for
the phase transition point $\Omega _{L}^{c}=0.4871E_{R}$. At $\Omega _{L}^{c}
$, the excitation spectrum contains two gapless Goldstone modes although the
anisotropic spin interaction is weak. $S(q_{x})$ has the very similar
feature as the spontaneous stripe phase, while $S_{\sigma }(q_{x})$ shows a
divergence at $q_{x}\rightarrow 0$. The effects of other external parameters
on $S(q_{x})$ and $S_{\sigma }(q_{x})$ are given in Fig.~\ref{FigureS4}. As
shown in Fig.~\ref{FigureS4}(a1)-(a3), $S(q_{x}=0)=0$, independent of
parameters. The dependence of $S(q_{x})$ on other parameters is generally
very weak.

In contrast, $S_{\sigma }(q_{x})$ shows strong dependence on other
parameters. It increases with increasing optical lattice strength $\Omega
_{L}$ (Fig.~\ref{FigureS4}(b1)), but decreases with increasing Raman
detuning $|\delta |$ [Fig.~\ref{FigureS4}(b2)]. $S_{\sigma }(q_{x})$ is
larger for weaker anisotropic spin interaction [Fig.~\ref{FigureS4}(b3)].
Interestingly, Fig.~\ref{FigureS4}(b3) shows that $S_{\sigma }(q_{x})$
increases (decreases) monotonically with the momentum for strong (weak)
anisotropic spin interaction, therefore has maximum (minimum) at $q_{x}=0$.
This result was also shown in Fig. 4 in the main text.

\end{document}